\documentclass[a4paper,10pt,twocolumn,aps,prl,nolongbibliography,superscriptaddress,showpacs,showkeys,amsmath,amssymb,floatfix,nofootinbib]{revtex4-1}

\usepackage{graphicx}
\usepackage{amssymb}

\usepackage[T1]{fontenc}
\usepackage[utf8]{inputenc}
\usepackage[dvipsnames]{xcolor}
\usepackage[unicode=true,pdfusetitle,
 bookmarks=true,bookmarksnumbered=true,bookmarksopen=true,bookmarksopenlevel=1,
 breaklinks=false,pdfborder={0 0 0},backref=false,colorlinks=true]
 {hyperref}
\hypersetup{
 citecolor=blue,filecolor=blue,linkcolor=blue,urlcolor=blue}
\makeatletter

\usepackage{textcomp}
\usepackage[caption=false]{subfig}
\usepackage{slashed}
\usepackage{braket}
\usepackage{graphicx}
\usepackage{float}
\pdfpageheight\paperheight
\pdfpagewidth\paperwidth

\renewcommand{\[}{\begin{equation}}
\renewcommand{\]}{\end{equation}} 

\usepackage{array}
\usepackage[normalem]{ulem}
\hypersetup{
 citecolor=blue,filecolor=blue,linkcolor=blue,urlcolor=blue}
\makeatother

\begin{document}

\title{Obtaining Precision Constraints on Modified Gravity with Helioseismology}

\author{Ippocratis D. Saltas} 
\email{ippocratis.saltas@fzu.cz}
\affiliation{CEICO, Institute of Physics of the Czech Academy of Sciences, Na Slovance 2, 182 21 Praha 8, Czechia}

\author{Il\'idio Lopes}
\email{ilidio.lopes@tecnico.ulisboa.pt}
\affiliation{Centro de Astrof\'{\i}sica e Gravita\c c\~ao  - CENTRA, Departamento de F\'{\i}sica, Instituto Superior T\'ecnico - IST, Universidade de Lisboa - UL, Av. Rovisco Pais 1, 1049-001 Lisboa}
\affiliation{Institut d'Astrophysique de Paris, UMR 7095 CNRS, Universit\'e Pierre et Marie Curie, 98 bis Boulevard Arago, Paris 75014, France}

\begin{abstract}
We propose helioseismology as a new, precision probe of fifth forces at astrophysical scales, and apply it on the most general scalar--tensor theories for dark energy, known as Degenerate Higher-Order Scalar-Tensor theories (DHOST). We explain how the effect of the fifth force on the solar interior leaves an observable imprint on the acoustic oscillations, and under certain assumptions we numerically compute the non-radial pulsation eigenfrequencies within modified gravity. We illustrate its constraining power by showing that helioseismic observations have the potential to improve constraints on the strength of the fifth force by more than $2$ orders of magnitude, as $-1.8 \cdot 10^{-3} \leq Y  \leq 1.2 \cdot 10^{-3}$ (at $2\sigma$). This in turn would suggest constraints of similar order for the theory's free functions around a cosmological background ($\alpha_{\text{H}}, \beta_{1}$). 
\end{abstract}
\maketitle

\section{Introduction}
General Relativity (GR) has been successful at describing observations at a vast range of scales, but its currently being challenged by crucial cosmological and astrophysical observations: the pressing questions of dark matter and dark energy suggest the possibility of new degrees of freedom and forces, yet to be discovered. 

The most popular extensions of GR are theories introducing a new dynamical scalar field coupled to spacetime. Intense efforts over the last years led to the remarkable construction of the most general, covariant theory describing the dynamics of a scalar field kinetically interacting with gravity, collectively labelled as DHOST scalar-tensor theories \cite{Zumalacarregui:2013pma,horndeski, BenAchour:2016fzp,Langlois:2017mxy, Langlois:2018dxi, Kobayashi:2019hrl}. They correspond to non-trivial generalisations of the popular Horndeski theory of gravity \cite{Horndeski:1974wa}, incorporating the archetypal Brans--Dicke theory as a subcase. Their cosmological and astrophysical phenomenology is rich, with their most notable prediction being the change in the propagation speed of gravitational waves as compared to GR. The recent measurement of the speed of tensors \cite{TheLIGOScientific:2017qsa, Monitor:2017mdv} placed the most stringent constraints on their theory space so far, ruling out a significant part of the allowed kinetic scalar-tensor interactions \cite{Ezquiaga:2017ekz,Creminelli:2017sry,Sakstein:2017xjx,Baker:2017hug, Dima:2017pwp, Kobayashi:2018xvr}. 
\\

An intriguing feature of the remaining non-trivial scalar-metric interactions is the prediction of a fifth-force effect within compact objects as \footnote{Technically, this is due to the breaking of the Vainshtein screening mechanism in the star, which would otherwise prevent sizable fifth-force effects.} \cite{Kobayashi:2014ida,Crisostomi:2017lbg,Dima:2017pwp},
\begin{align}
\nabla^2 \Phi = 4 \pi G \rho + G \frac{Y}{4} \nabla^2 \left( \frac{dm}{dr} \right), \label{Poisson}
\end{align}
with $\Phi$ the gravitational potential, $m(r)$ the mass enclosed within radius $r$, and $Y$ the coupling strength of the new force. A $Y > 0$ ($Y<0$) tends to weaken (strengthen) gravity, since $d \rho/dr < 0$ in the stellar interior, while Newtonian gravity is recovered outside the star ($dM/dr \to 0$).
In a cosmological context, $Y$ relates to the parameters associated with the large-scale structure dynamics of general scalar-tensor theories labelled as $\alpha_{\text{H}}$ and $\beta_{1}$ \cite{Dima:2017pwp}. Therefore, constraints on $Y$ have direct consequences for gravity at large scales and dark energy modelling. Currently, the upper and lower bound from astrophysics comes from white dwarfs as $Y > -0.48$ \cite{Babichev:2016jom} and $Y < 0.18$ \cite{Saltas:2018mxc} respectively (see also Refs. \cite{Sakstein:2018fwz, Ishak:2018his}). 
\\

Our goal is to explore helioseismology as a high-precision test of fifth forces at local scales \footnote{For detailed expositions on helioseismology see Refs. \cite{Gough-lectures, Aertsbook, Unno, ChristensenDalsgaard:2002ur}.}, focusing on theory (\ref{Poisson}). 
The solar eigenspectrum traces the finest details of the solar interior, which in combination with the accuracy of the observed frequencies ($\sim 1$ in $10^5$), provides a powerful probe of the underlying physics. We will explain how the fifth force leaves an observable imprint on the solar eigenspectrum through the subtle deformations of the solar sound speed profile, and employing helioseismic simulations we will illustrate the power of helioseismological constraints in this regard. 

\section{Helioseismology as a powerful probe of gravity \label{sec:Theory}} 

For an intuitive grasp of the way helioseismology traces the interior solar physics, let us consider a key result of the asymptotic (WKB) theory of stellar pulsations, describing the characteristic frequency of an acoustic wave associated to the travel time from the stellar center to the surface (see e.g Refs. \cite{Aertsbook, Gough-lectures}), 
\begin{equation}
f_{\text{acoustic}} = \left(\frac{1}{2} \int_{0}^{R} \frac{dr}{c_{s}} \right)^{-1}, \label{f-asymptotic}
\end{equation}
with $R$ the surface radius and $c_{s} \equiv \sqrt{\partial P/\partial \rho}$ the interior speed of sound, while $P$ and $\rho$ stand for the pressure and density profiles. Clearly, solar pulsations probe not only global properties of the star, but also the structure of its interior medium through the shape of the sound speed. 
Within a helioseismic inversion context, this discrepancy is translated to corrections on physical interior profiles and microphysics assumptions, which can be elegantly formulated as a variational principle (see e.g Refs. \cite{ChristensenDalsgaard:2002ur, Gough-inversion}).

\vspace{0.1cm}

Stellar non-radial adiabatic pulsations correspond to small departures from the spherically symmetric equilibrium state, described by a system of fourth-order differential equations for the displacement vector, Lagrangian pressure, and Eulerian gravitational potential field perturbations. Defining ${\bf \delta X} = \{\delta {\bf r}, \delta P, \delta \Phi, d \delta \Phi/dr\}$ \footnote{We consider $\delta P$ and $\delta \Phi$ as a Lagrangian and Eulerian perturbation respectively.}, we write
$
\mathcal{L}[P(r), c_{s}(r);  {\bf \delta X} ] = 0, 
$  
with $\mathcal{L}$  a linear differential operator \cite{Unno}. The eigenspectrum is computed requiring regularity at the center, and that the pressure perturbation vanishes at the surface (free-boundary condition \cite{Unno}). In a spherical harmonic basis, modes with $l \neq 0$ ($l =0$) correspond to non-radial (radial) ones, while the overtone $n$ counts the number of radial nodes. Solutions are standing waves formed in the cavity defined by an interior turning point, $r_\text{t}$, and an external turning point. The interior turning point shifts out to the surface for the higher-degree acoustic modes.
\vspace{0.1cm}
\\

\begin{figure*}
\hspace{-2.5cm} \centering
\begin{minipage}[b]{.55\textwidth}
\includegraphics[scale=0.4]{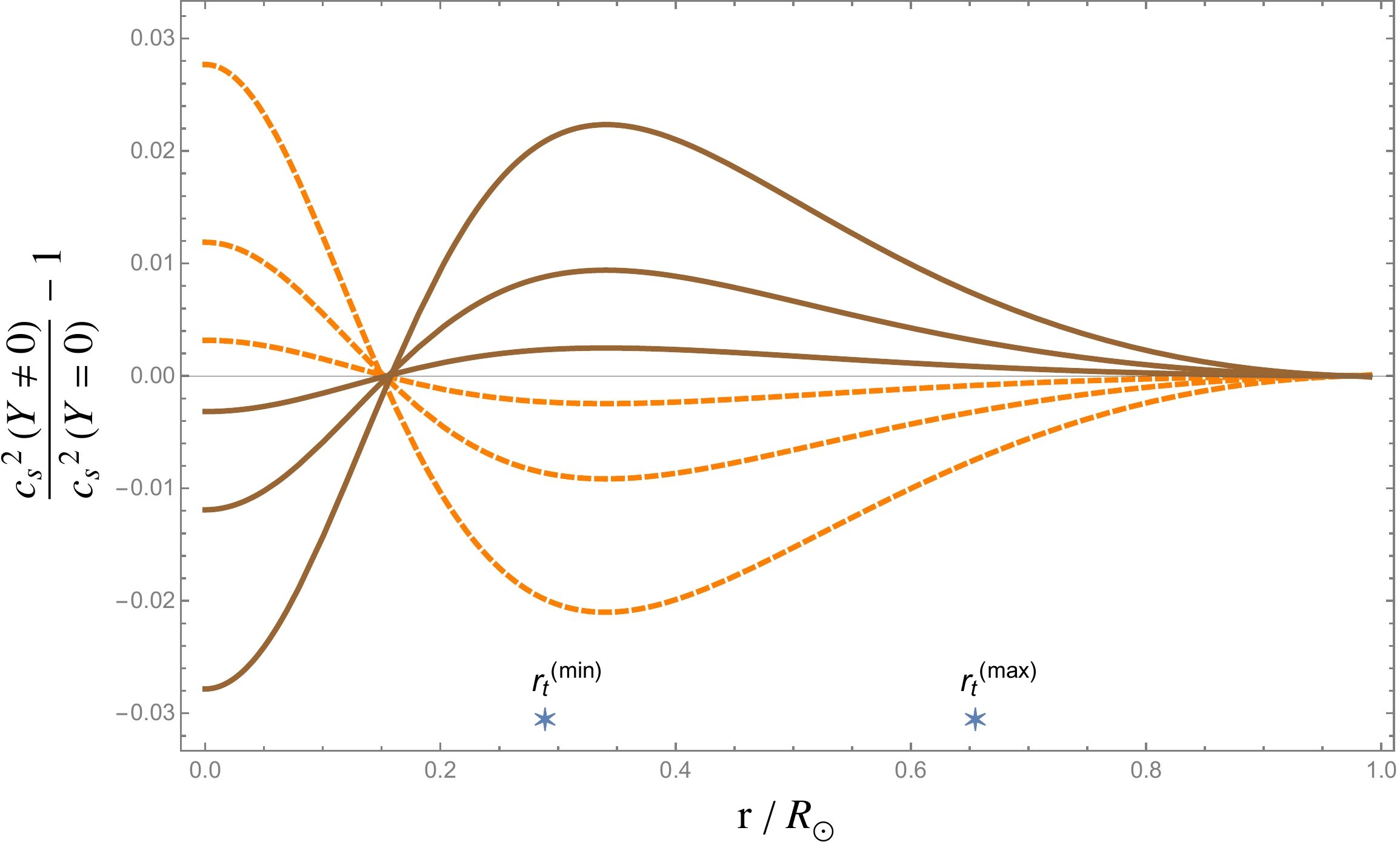}
\end{minipage}
\begin{minipage}[b]{.44\textwidth}
\includegraphics[scale=0.36]{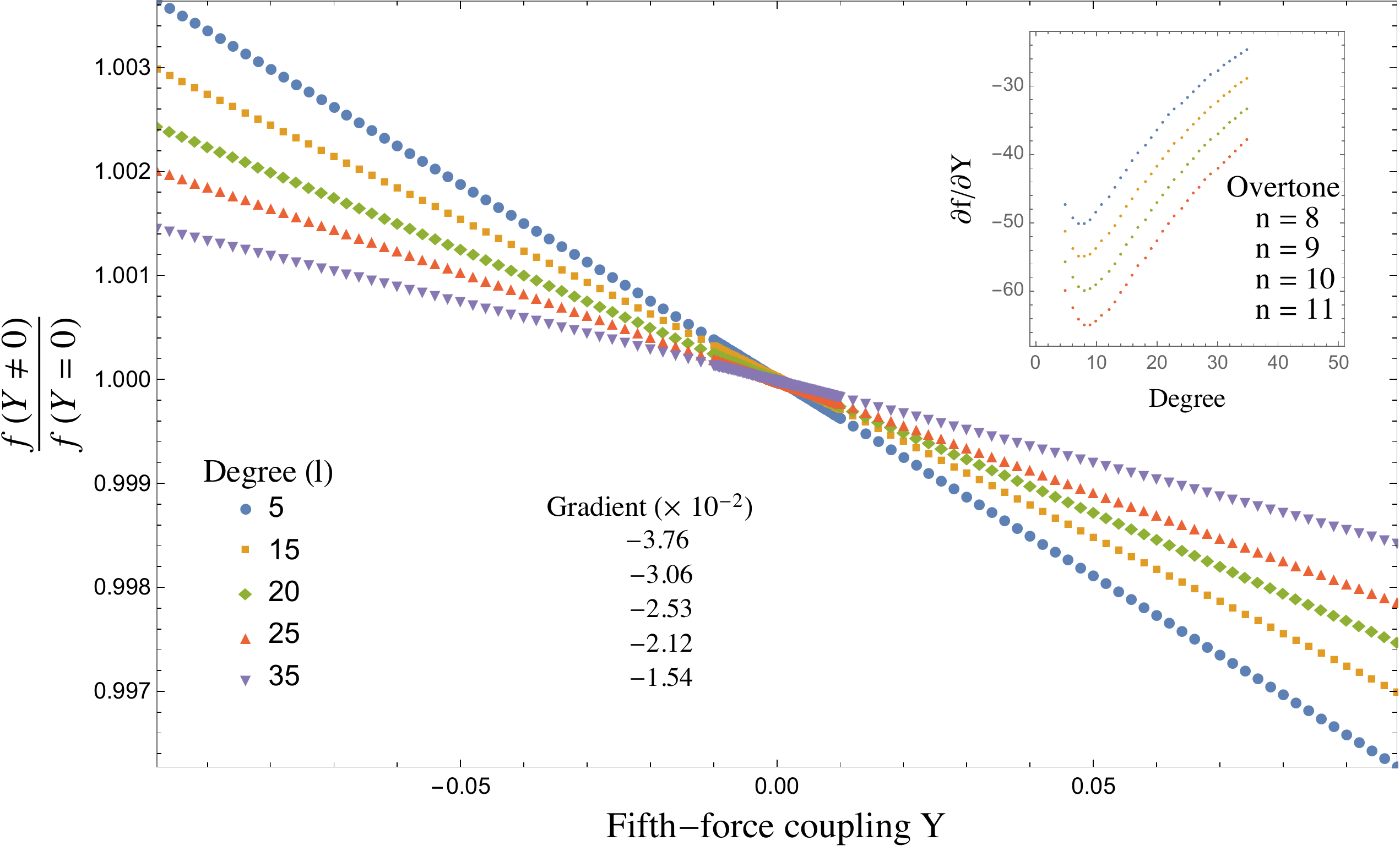} 
\end{minipage}
\caption{{\bf Left}: Fractional difference of the solar sound speed between Newtonian and modified gravity for the indicative values of the fifth-force coupling $Y = \{ \pm \, 7 \cdot 10^{-2}, \pm \, 3 \cdot 10^{-2}, \pm \, 8 \cdot 10^{-3} \}$, based on the polytrope. Continuous (dashed) curves correspond to weaker (stronger) gravity with $Y > 0$ ($Y < 0$). 
The typical range of turning points of the modes considered are shown with $r_{t}^{(\text{min})} =0.309  \, R_{\odot}$ and $r_{t}^{(\text{max})} =0.676 \, R_{\odot}$, corresponding to $l=5$ and $l=35$ at $n =10$.
{\bf Right}: Scaling of polytropic frequencies with $Y$ at fixed overtone $n = 10$.
An order $10 \%$ weakening or strengthening of gravity ($|Y| \sim 0.1$) induces a $\sim 0.1 \%$ change in the frequencies.
The inset shows the dependence of $\partial f/\partial Y$ on overtone and degree. Larger degrees probe outer solar regions where the fifth-force effect decreases, leading to a decreasing $|\partial f/\partial Y|$.}  
\label{fig:cs}
\end{figure*}

{\it Modified gravity}: 
The modified Poisson equation (\ref{Poisson}) implies a new hydrostatic equilibrium as
\begin{align}
\frac{dP}{dr} = - \frac{GM(r)}{r^2}\rho(r) -  \frac{G \cdot Y}{4}\frac{ d^{2} M(r)}{d r^2} \rho(r), \label{hydro-eq}
\end{align}
which in turn implies a new pulsation spectrum due to the modified equilibrium structure of the star.
Fig. \ref{fig:cs} (left) shows the fractional change in $c_{s}$, $\Delta c_{s}^{2}/c_{s}^{2}$, under the theory (\ref{Poisson}), based on a polytrope. Each region of the star is impacted differently, with the effect escalated at two regions: the center, and an interior point ($r \sim 0.3 R_{\odot}$). The effect becomes stronger with $|Y|$, while a weakening (enhancement) of gravity tends to shift the interior peak towards (away from) the center. 
In turn, the modified sound-speed profile impacts on the predicted acoustic eigenspectrum. The right of Fig. \ref{fig:cs} shows the scaling of eigenfrequencies with $Y$, numerically computed according to the procedure to be explained later. Frequencies become smaller (larger) for weaker (stronger) gravity, as qualitatively expected considering Eq. (\ref{f-asymptotic}) in combination with the response of $c_{s}$ to the fifth force. The effect of $Y$ becomes more pronounced for smaller degrees $l$ (at fixed $n$), as reflected by the larger slope, as modes with larger degrees probe outer parts of the star, where $\Delta c_{s}^2/c_{s}^2$ is declining. 
\\ \vspace{0.1cm}

{\it Cowling approximation}: A highly useful approximation widely utilised in asteroseismological studies is the Cowling approximation \cite{Cowling}, which accounts to neglecting the Eulerian perturbation of the gravitational potential $\delta \Phi$ for sufficiently large degrees $l \gg 1$, due to its overall suppression by the large factor $\sim \frac{1}{(2l + 1)}$. Therefore, the backreaction of gravity to the density perturbation is not accounted for, and the information about gravity enters implicitly through the background configuration. Since the fifth force acts as a perturbative correction to the Newtonian term, for the perturbed potential, $\delta \Phi = \delta \Phi_{\text{N}} + \delta \Phi_{\text{MG}}$, this translates to $|\delta \Phi_{\text{MG}}| \ll |\delta \Phi_{\text{N}}|$. We will therefore apply the approximation to the full potential, expected to hold except for possible singular, unphysical configurations. This will allow for an insightful understanding of the problem without loss of accuracy -- a study of the full non-radial oscillation equations in modified gravity goes beyond our current scope.

\begin{figure}
\hspace{-1.5cm} \includegraphics[width=.5\textwidth]{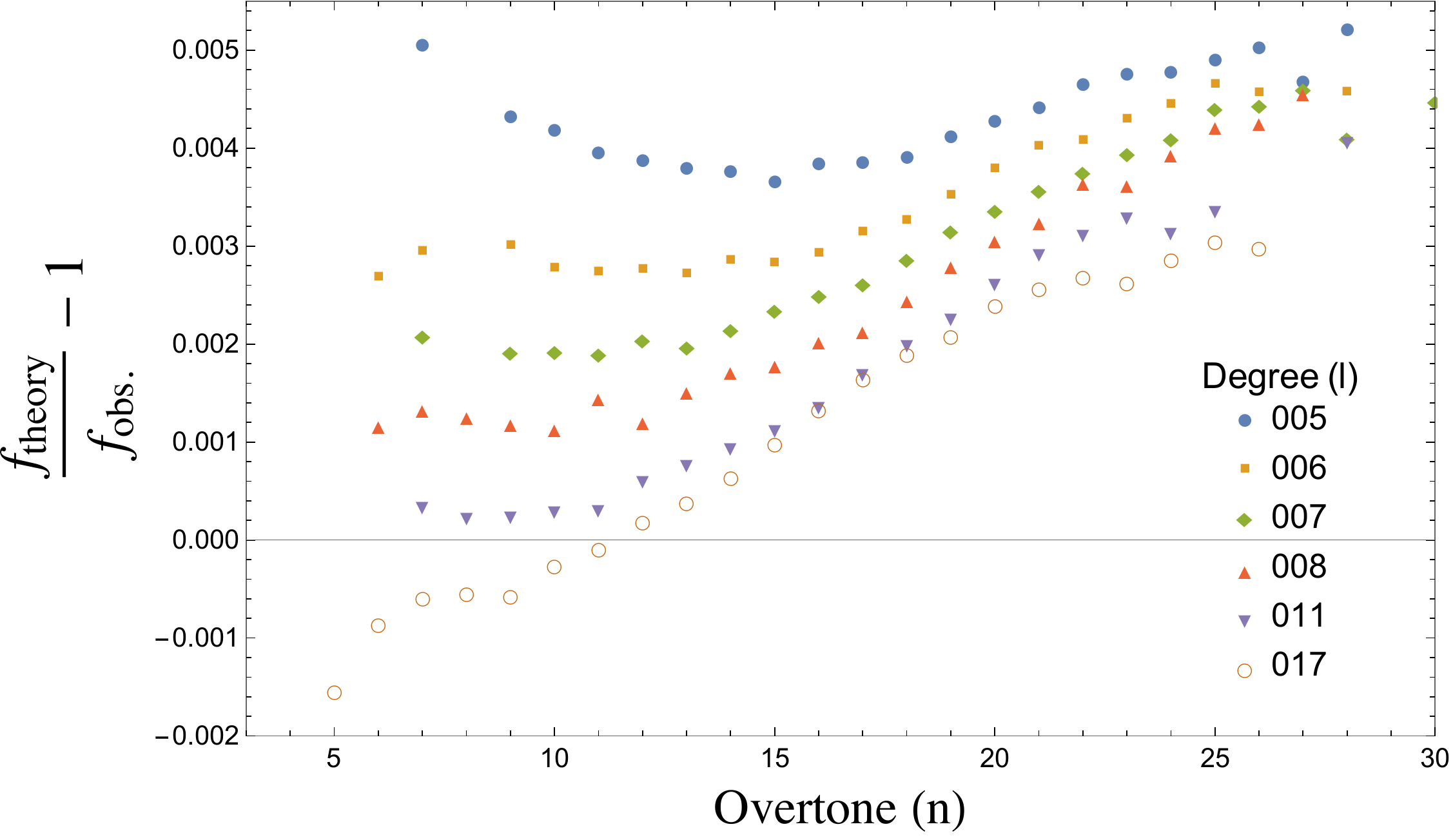}
\caption{Fractional difference between the predicted and observed frequencies for selected modes, based on our reference evolutionary model at standard gravity, computed under the Cowling approximation and boundary conditions explained in the text. Missing overtones are due to the unavailability of observational data for those modes. Given a reference stellar model, the discrepancy between the predicted and observed frequencies is typically a signal of inadequate modelling in the star's interior physics, and the underlying idea of helioseismic inversions. 
}\label{plot:frequency-overtone}
\end{figure}

\section{Modelling and computation of the pulsation frequencies \label{sec:simulations}} 

{\it Eigenspectrum computation:} We use the state-of-the-art oscillation suite GYRE \cite{Townsend:2013lua}, a modular code combining advanced shooting/integration schemes to accurately solve the boundary value problem of the pulsation equations on a spatial grid. For the outer boundary condition, we impose that the pressure variation vanishes at $r = R_{\odot}$ ($\delta P = 0$). For a detailed description of the equations see Refs. \cite{Townsend:2013lua, Unno, Dziembowski}. 
\\ \vspace{0.1cm}

{\it Data set:} We will utilise the helioseismic data of the {\it Global Oscillation Network Group (GONG)}  \cite{GONG}, with measurements of solar eigenfrequencies between $l = 0 -120$, at the accuracy of $1$ part in $ 10^{4}$. 
\\ \vspace{0.1cm}

{\it Solar modelling and eigenspectrum:} The interior modelling of the star goes hand in hand with the predicted eigenspectrum. 
In this regard, we first use the code MESA \cite{Paxton2011,Paxton2013,Paxton2015} to produce an evolved model of the present Sun assuming standard gravity. We evolve a $1M_{\odot}$ star from its zero-age main sequence, calibrated so that for the solar radius $R/R_{\odot} \sim 10^{-4}$, luminosity $L/L_{\odot} \sim 10^{-3}$, convective radius $R^{cz}/R^{cz}_{\odot} \sim 10^{-1}$ and rms sound speed $|c_{s}^{\text{rms}} - c_{s \odot}^{\text{rms}}| \sim 10^{-4}$, after tuning the element abundances, mixing length/overshooting parameters, and choosing the Krishna Swamy atmospheric model. The equation of state used by the code MESA relies on a combination of different equations of state, in particular, the so--called OPAL \cite{Rogers2002}, SCVH\cite{Saumon1995}, PTEH \cite{Pols1995}, HELM \cite{Timmes2000}, and PC \cite{Potekhin2010}. The bulk of opacities come from OPAL \cite{Iglesias1993,
Iglesias1996}, with their low- and high-temperature description from \cite{Ferguson2005} and \cite{Buchler1976} respectively. Electron-conduction opacities are based on \cite{Cassisi2007}. Nuclear and and weak reaction rates are from JINA REACLIB \cite{Cyburt2010}  and \cite{Fuller1985, Oda1994, Langanke2000} respectively. 

We proceed computing the acoustic eigenspectrum based on the evolutionary model ($Y = 0$) with GYRE, considering $5 \leq l \leq 35$, and scanning for frequencies up to the $n \sim 40$th overtone. The choice of $l \geq 5$ is for consistency with the Cowling approximation. The numerical computation reveals that agreement between predicted and observed frequencies is at the $0.1 \%$ level, see also Fig. \ref{plot:frequency-overtone}.
\\ \vspace{0.1cm}

{\it Modified gravity:}  Aiming at the best balance between simplicity and accuracy, we proceed in two steps: First, we compute the eigenspectrum in modified gravity using as our proxy a polytropic equation of state. The polytrope inevitably omits for a variety of microphysics -- this offset is then compensated for at the level of the eigenfrequencies in an effective manner, using the results of our reference evolutionary model at standard gravity ($Y = 0$).

For the polytropic index, we fix $n_{\text{pol}}  =3.069$, which we find to provide the best-fit to the density and pressure profiles of the evolutionary model, so that consistency between both descriptions is ensured. 
We first construct a set of polytropic models based on Eq. (\ref{hydro-eq}) for $- 10^{-1} \leq |Y| \leq 10^{-1}$ and step size $\delta Y = 0.2 \cdot 10^{x}$, $x \in \{ -6, \ldots, -1 \}$, and proceed solving the pulsation equations with GYRE on a spatial grid of $\, \sim 7 \cdot 10^4$ points and $5 \leq l \leq 35$. The resulting dependence of frequencies on $Y$ is shown in Fig. \ref{fig:cs}.

Comparison between the polytropic (at $Y = 0$) and evolutionary-model frequencies suggests they disagree by $17.5 \% - 21.5 \%$.
We compensate for this offset effectively, through a correction term $\delta f$, which accounts for all corrections from a more accurate accounting of the microphysics as
\begin{equation}
f_{\text{theory}} (n, l; Y) = f_{\text{pol.}}(n, l; Y) + \delta f(n,l; Y). \label{freq-corrected}
\end{equation}
$f_{\text{theory}}$ denotes the predicted and {\it sufficiently accurate} frequency, while the polytropic $f_{\text{pol.}}$ is constructed with $n_{\text{pol}} = 3.068$. Since the fifth force is a perturbative correction to the Newtonian term, we can estimate $\delta f(n,l; Y)$ through an expansion around $Y = 0$ as 
$
\delta f \simeq \delta f_{(0)} + \delta f_{(1)} \equiv \delta f_{(0)} + \left. \frac{\partial \delta f(Y)}{\partial Y}\right|_{Y = 0}\cdot Y.
$
The term $\delta f_{(0)}$ accounts for the neglected microphysics of the polytrope at $Y = 0$, while $\delta f_{(1)}$ measures their response to the fifth force. Therefore, for $Y = 0$, $f_{\text{theory}}$ identifies with the frequencies computed with the evolutionary model, allowing us to explicitly extract the zeroth-order correction as $\delta f_{(0)} = f_{\text{theory}} - f_{\text{pol.}}(Y=0)$. 

Now, it is straightforward to see that the linear correction $\delta f_{(1)}$ can be written in the suggestive form 
\begin{align}
\delta f_{(1)} \equiv  f_{\text{pol.}}'\, \xi  \cdot Y, \; \;  \;\xi \equiv \left( \frac{f_{\text{pol.}}'  - f_{\text{theory}}'}{f_{\text{pol.}}'} \right)_{Y = 0}, \label{lin-term0}
\end{align}
with derivatives evaluated at $Y = 0$, gradients depending on ($n, l$), and $\partial f/\partial Y \equiv f'$. In the absence of evolutionary simulations in modified gravity, $f_{\text{theory}}'$ is unknown, and we will instead treat $\xi$ as a nuisance parameter. Notice that, since gradients depend on $(l, n)$, Eq. (\ref{lin-term0}) accounts for the appropriate weight for each mode through $f_{\text{pol.}}'$, the value of which is known from our numerical simulations. Our goal now is to derive an upper bound for $|\xi|$ that will guide its marginalisation range later on. 
For a numerical estimate, we consider the approximate result from the asymptotic (WKB) analysis for sufficiently low-degree modes (see, e.g., Refs. \cite{Aertsbook, Gough-lectures}),
\begin{align}
f = \left( n + \frac{l}{2} + \alpha \right) \bar{f}, \label{f-asymptotic2}
\end{align} 
with $\bar{f} \equiv f_{\text{acoustic}}$ given in Eq. (\ref{f-asymptotic}), and $\alpha$ a correction due to the phase shift as the wave is reflected at the outer boundary. It is $\alpha \simeq n_{\text{pol}}/2$ for a purely polytropic star, and $\alpha \simeq 0.646$ for perfectly conducting atmospheres (see, e.g., Refs. \cite{Aertsbook, Gough-lectures} for details).

To estimate $f_{\text{theory}}'(Y)$ in Eq. (\ref{lin-term0}), we first differentiate (\ref{f-asymptotic2}) with respect to $Y$. In turn, this requires an estimate of the corrected, fundamental frequency $\bar{f}_{\text{theory}}(Y)$. To compute it, we use Eq. (\ref{f-asymptotic}) under a similar improvement to Eq. (\ref{freq-corrected}), but at the level of the sound speed as $c_{s}(Y) = c_{s}^{\text{pol.}}(Y) + \delta c_{s}(Y) \simeq c_{s}^{\text{pol.}}(Y) +  \delta c_{s}^{(0)}$, truncated at zeroth order \footnote{It is straightforward to see that, expanding the integrand in (\ref{f-asymptotic}) with respect to $Y$, the linear correction $\delta c_{s}^{(1)}(Y)$ is suppressed by at least one order of magnitude compared to the zeroth-order term.}.
The polytropic piece acts as a proxy to the fifth-force's effect , while the $Y-$independent correction $\delta c_{s}^{(0)}$ is extracted comparing the sound speed profile of our evolutionary model with the polytrope (at $Y = 0$). Evaluating Eq. (\ref{f-asymptotic}) under this approximation, and differentiating the result leads to $\partial \bar{f}_{\text{theory}}/\partial Y \simeq - 3.425$. For the pure polytrope ($n_{\text{pol}} = 3.069$), it is easy to see that $\partial \bar{f}_{\text{pol.}}/\partial Y \simeq - 3.053$. We also use $\alpha_{\text{pol.}} = n_{\text{pol}}/2$ and $\alpha_{\text{theory}} = 0.646$ respectively. Now, our numerical solutions for $f_{\text{pol.}}' $ show that its magnitude increases with increasing degree at fixed overtone, approximately according to (\ref{f-asymptotic2}) up to $l \sim 10$, and starts decreasing with $l$ beyond that point (see right of Fig. \ref{fig:cs}). Therefore, using Eq. (\ref{f-asymptotic2}) and the previous estimates, an upper bound is found substituting the highest overtone in the particular subsample of modes we use for our statistical analysis (see next section), $n = 11$, for which $l_{\text{critical}} = 9$, yielding $|\xi|_{\text{upper}} \simeq 5.2 \%$. We allowed for both positive and negative values in $\xi$ for consistency, given that its exact value is unknown in our analysis.

\section{The constraining power of helioseismology on the fifth force coupling and its cosmological implications \label{sec:obs}} 
Typically, stellar-evolutionary models cannot accurately enough predict the star’s eigenspectrum. Within helioseismic inversions, statistically significant differences between theory and observations are translated to background-modelling corrections. In this context, disentangling the subtle effects of the fifth force from such systematics proves a challenging task. A helioseismic-inversion analysis in modified gravity goes beyond the scope of this work -- instead, to minimise background-modelling systematics, we select those frequencies differing by no more than $1 \sigma_{\text{obs.}}$, i.e $|f_{\text{theory}} - f_{\text{obs}}| < \sigma_{\text{obs.}}$, when computed with the evolutionary model (at $Y = 0$). We find this is satisfied in total by $19$ modes ranging between $l = 10 - 35$ and $n = 6 -11$.  
For all modes of this subset we construct a combined likelihood as $\mathcal{L}(Y; \xi) \propto \text{exp}\left( - \chi^{2}/2 \right)$, with 
$
\chi^{2}(Y, \xi; \, l,n) \equiv \sum_{l, n} \Big[f_{\text{theory}}(Y, \xi; \, l,n) - f_{\text{observed}}(l,n) \Big]^2/\sigma_{\text{obs.}}^2,
$
and  $f_{\text{theory}}$ computed according to the previous section. 
\\

Using $|\xi|_{\text{upper}} \simeq 5.2 \%$, we first marginalise $\mathcal{L}$ over $\xi \in [-0.052, 0.052]$, to find that, $-1.71 \cdot 10^{-3} \leq Y  \leq 1.15 \cdot 10^{-3}$ ($2\sigma$). Had we naively assumed the fractional error on the frequency gradients is similar to that for the frequencies between the polytropic and MESA model ($\sim 17.5 - 21.5 \%$), we are led to the extreme case of $|\xi| \simeq 22$. Marginalising over $|\xi| \leq 22$ we find $-1.79 \cdot 10^{-3} \leq Y  \leq 1.2 \cdot 10^{-3}$ ($2\sigma$) --  Clearly, the marginalisation range has no practical effect, and using the latter conservative choice for $|\xi|$, we can quote at $2\sigma$
\begin{align}
-1.8 \cdot 10^{-3} \leq Y  \leq 1.2 \cdot 10^{-3}. \label{Y-constraint}
\end{align}
This suggests an improvement by more than $2$ orders of magnitude on the previous lower ($Y > -0.48$ \cite{Babichev:2016jom}) and upper ($Y < 0.14$ \cite{Saltas:2018mxc}) bounds from astrophysics, and adds to intense previous efforts to constrain $Y$ \cite{Jain:2015edg, Babichev:2016jom, Saltas:2018mxc,Sakstein:2015zoa,Sakstein:2015aac,Sakstein:2018fwz, Sakstein:2016ggl,Babichev:2018rfj,Creminelli:2018xsv,Crisostomi:2019yfo}.
The result (\ref{Y-constraint}) should be understood within the context of our data-selection criterion, relying on the modes best described by our reference model. We find that, inclusion of a broader set of modes causes tension with Newtonian gravity at $2\sigma$ \footnote{This is due to a shift in the total likelihood's central value, and not a change in its spread.}, until the point when individual likelihoods are in tension with each other too -- the latter tension prevents the extraction of a global constraint on $Y$ from all data points, and indicates the need for improved modelling. A detailed helioseismic-inversion treatment in the future would allow for a consistent analysis of all modes, and the distinction between genuine fifth-force effects from background-modelling artifacts. Therefore, Eq. (\ref{Y-constraint}) illustrates the constraining power of our approach, and should be regarded as a first, order-of-magnitude estimate of what would be a thorough helioseismic analysis. At the same time, one might wonder how sensitive our results are if we instead impose a stricter condition. For example, requiring the difference between predicted frequencies at standard gravity and observed ones is no more than $\sigma_{\text{obs.}}/2$ or $\sigma_{\text{obs.}}/4$, the modes available reduce to $14$ and $4$ respectively. For the extreme case of $4$ data points, we find $-2.52 \cdot 10^{-3} \leq Y  \leq 2.78 \cdot 10^{-3}$ ($2\sigma$) -- Although not as robust as the actual sample, it provides with a measure of the constraint's dependence on our data-selection criterion. 
\\

\begin{figure} 
     \includegraphics[scale=0.4]{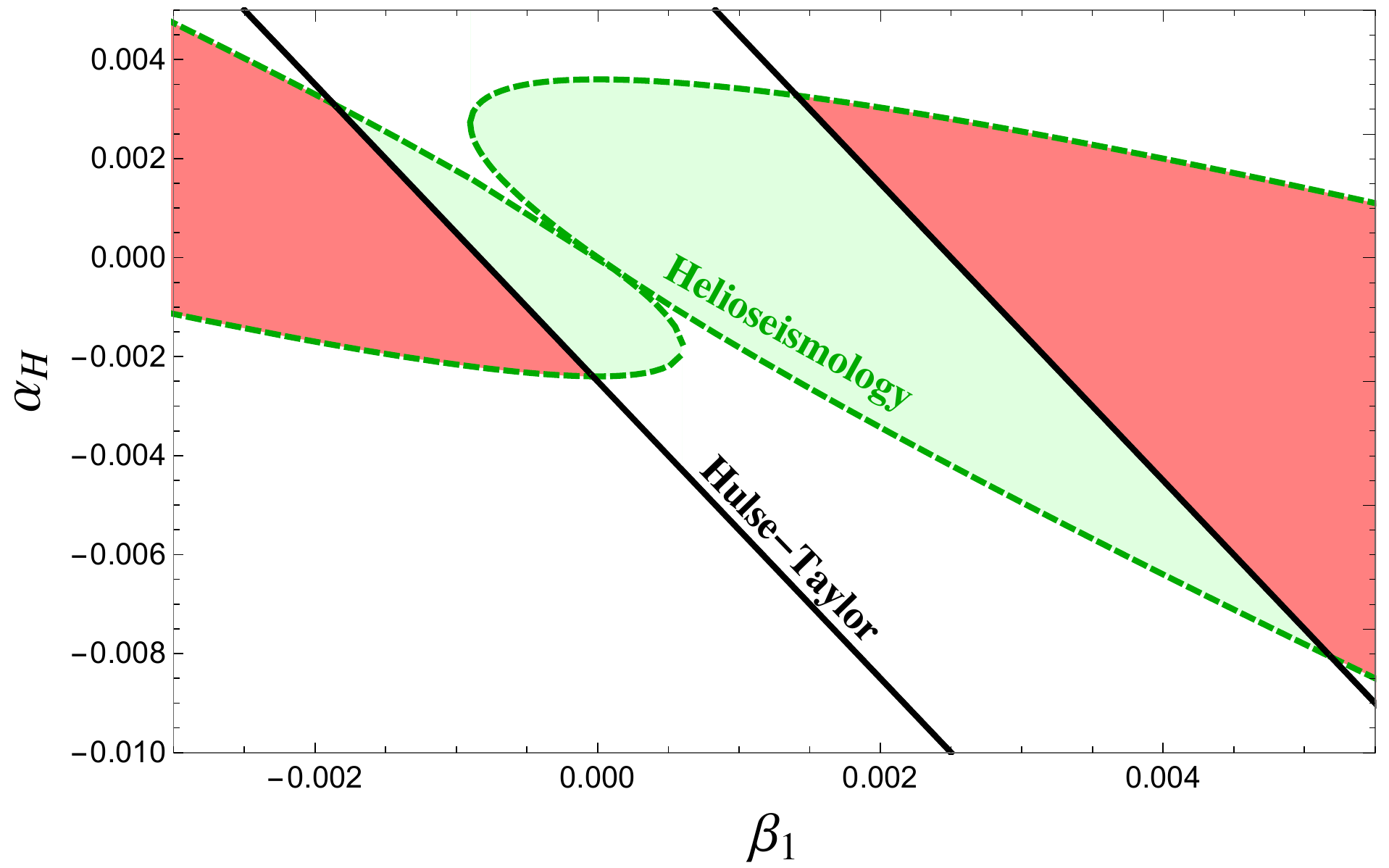}
      \caption{The implications of Eq. (\ref{Y-constraint}) on the cosmological-parameter space $(\beta_{1}, \alpha_{\text{H}})$. The region between the black-continuous boundaries corresponds to the constraint of \cite{Dima:2017pwp}, while green-dashed ellipses to the predicted helioseismological bound (\ref{Y-constraint}).
                 }   \label{fig:alphabeta}
\end{figure}

The coupling $Y$ relates to effective theory functions of the original scalar-tensor theory as
$
Y = - \frac{2 (\alpha_{\text{H}} + \beta_1)^2}{\alpha_{\text{H}} + 2\beta_1}
$ \cite{Dima:2017pwp}. 
$\beta_{1}(t)$ parametrises the contribution of higher-order scalar-metric kinetic operators in the action, while $\alpha_{\text{H}}(t)$ quantifies the amount of kinetic mixing between the scalar and matter. Their current constraint is due to the Hulse-Taylor pulsar combined with white-dwarf observations, $-8 \cdot 10^{-2} \leq \beta_1  \leq 2 \cdot 10^{-2}$ and $-5 \cdot 10^{-2} \leq \alpha_{\text{H}}  \leq 2.6 \cdot 10^{-1}$ \cite{Dima:2017pwp}, while cosmological probes suggest $\mathcal{O}(1)$ constraints \cite{Traykova:2019oyx}. The result (\ref{Y-constraint}) implies for both parameters at $2\sigma$ as $ - 1.9 \cdot 10^{-3} \leq \beta_1  \leq 5.2  \cdot 10^{-3}$ and $ - 2.4 \cdot 10^{-3} \leq \alpha_{\text{H}}  \leq 3.3 \cdot 10^{-3}$. The implications of (\ref{Y-constraint}) on the plane of $(\beta_{1}, \alpha_{\text{H}})$ is shown in Fig. \ref{fig:alphabeta}, and it is to be compared with the similar figure of Ref. \cite{Dima:2017pwp}.

\begin{figure}
\includegraphics[width=.35\textwidth]{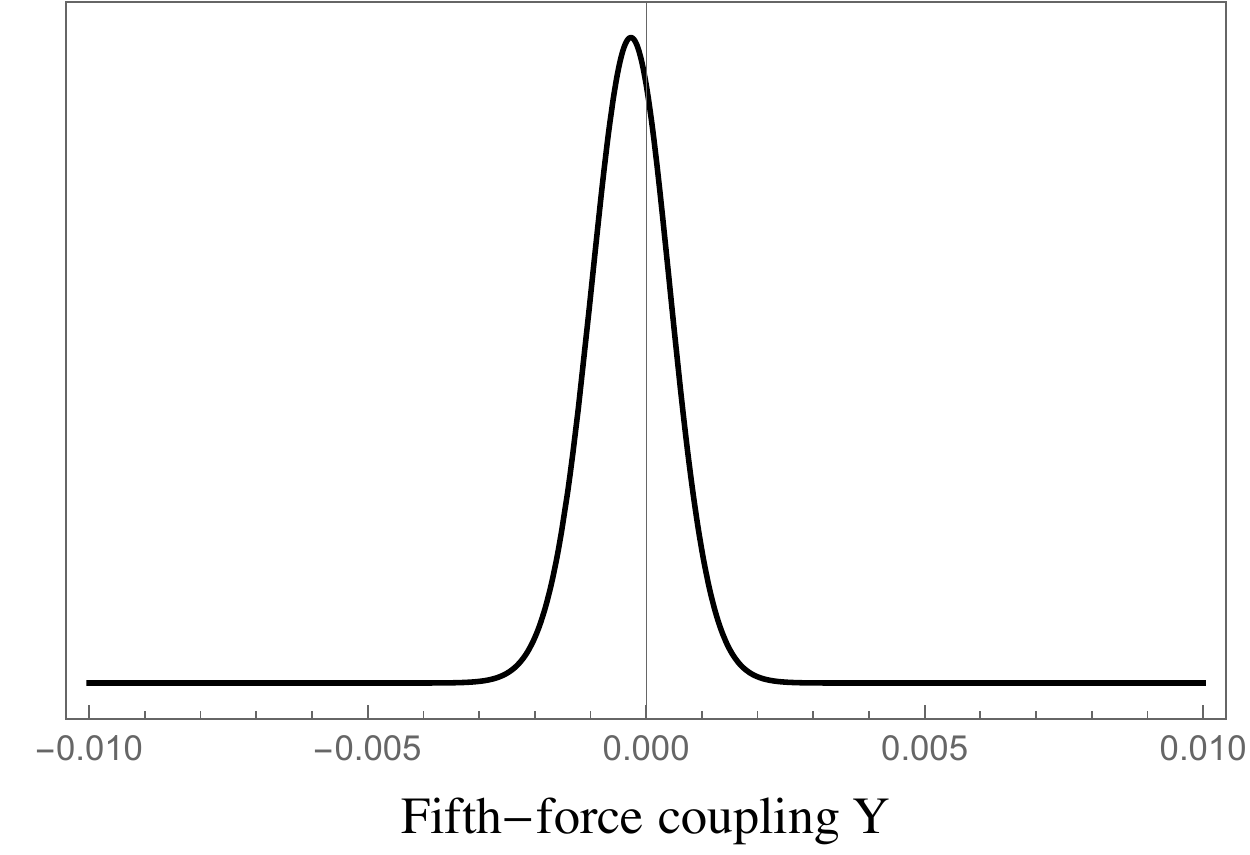}
\caption{The probability density leading to (\ref{Y-constraint}), $-1.8 \cdot 10^{-3} \leq Y  \leq 1.2 \cdot 10^{-3}$ ($2\sigma$). It results combining the $19$ selected pulsation modes, and marginalising over the $\xi$ parameter under a flat prior with $|\xi| \leq 0.22$, as explained in the text. 
}\label{plot:probdens}
\end{figure}

\section{Summary}
We proposed helioseismology as a high-precision test for the most general scalar-tensor theories (DHOST). We showed how the subtle fifth-force effect leaves a characteristic observable imprint on solar pulsations, and demonstrated the constraining power of our approach for the fifth-force coupling strength. 

This is the first step towards a complete treatment of helioseismology in modified gravity, begging for further studies in the search of new exciting effects, and the confirmation of our results with a fully consistent approach. In particular, going beyond the Cowling approximation and the inclusion of helioseismic corrections will allow us to probe a broad part of the eigenspectrum. In turn, this calls for the construction of the modified non-linear pulsations equations, along with accurate solar models in the presence of the fifth force. We leave these issues for future work.
\\

\begin{acknowledgments}

{\it Acknowledgements --} I.D.S. is grateful to J\o rgen Christensen-Dalsgaard for constructive criticism, feedback and fruitful discussions. We also thank Stefan Ilic and Ignacy Sawicki for useful discussions and feedback, as well as the anonymous referees for constructive criticism. Parts of the numerical simulations of this work were performed with the Koios Slurm cluster of the Czech Academy of Sciences. I.D.S. is grateful to Joseph Dvoracek for his valuable help with parallel computing and bash scripting. I.D.S. is funded by European Structural and Investment Funds and the Czech Ministry of Education, Youth and Sports (Project CoGraDS --- CZ.02.1.01/0.0/0.0/15\_003/0000437). 

\end{acknowledgments}

\bibliographystyle{utcaps}
\bibliography{UM.bib,WD-Bib.bib}

\end{document}